\documentclass[
    ,final            
  ]
  {aipproc}

\usepackage{wasysym} 

\layoutstyle{8x11double}
 
\newcommand{\hess}{H.E.S.S.}  
\newcommand{\hgc}{HESS~J1745-290}
\newcommand{\astar}{Sgr~A$^*$} 
\newcommand{\aeast}{Sgr~A~East}
\newcommand{\pwn}{G359.95-0.04}
\newcommand{\gr}{$\gamma$-ray}
\newcommand{\grs}{$\gamma$-rays}

\newcommand{\gsnr}{G~0.9+0.1}

\begin{document}

\title{Very High Energy $\gamma$-ray Observations of the Galactic Centre Region}


\author{Christopher van Eldik}{
  address={Max-Planck-Institut f\"ur Kernphysik, P.O. Box 103980, D-69029 Heidelberg, Germany}}

\begin{abstract}
Recent progress in pushing the sensitivity of the Imaging Atmospheric
Cherenkov Technique into the 10~mCrab regime has enabled first
sensitive observations of the innermost few 100~pc of the Milky Way in
Very High Energy (VHE; $>100$~GeV) \grs. These observations are a valuable tool to understand
the acceleration and propagation of energetic particles near the
Galactic Centre. Remarkably, besides two compact \gr\ sources,
faint diffuse \gr\ emission has been discovered with high
significance. The current VHE \gr\ view of the Galactic Centre
region is reviewed, and possible counterparts of the \gr\ sources
and the origin of the diffuse emission are discussed. The future
prospects for VHE Galactic Centre observations are discussed based on
order-of-magnitude estimates for a CTA type array of telescopes. 
\end{abstract}

\maketitle


\section{The inner few 100~pc of the Milky Way}
Ever since the discovery of the strong compact radio source \astar\
\cite{Balick:1974aa}, the Galactic Centre (GC)
has been subject to intense astrophysics and astronomy research. In
the last decade, precise data from this peculiar region, that evades
observations at optical wavelengths due to obscuration by dust along
the line-of-sight, have been obtained at radio, infrared (IR), X-ray, and
hard X-ray/soft \gr\ energies. Because of its proximity, the GC is a
unique, however complex, laboratory for investigating the
astrophysics believed to be taking place in galactic nuclei in general. 

\begin{figure}
\includegraphics[width=0.48\textwidth]{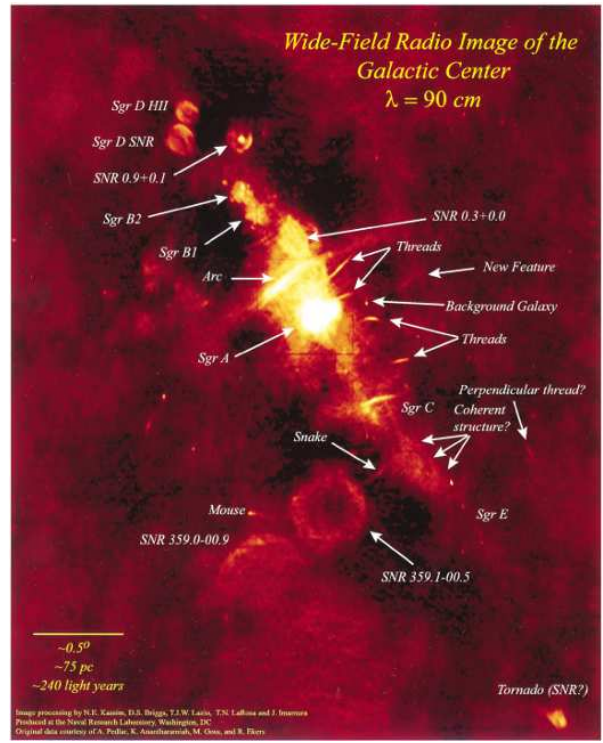}
\label{fig:LaRosa}
\caption{Large-scale compilation of VLA 90~cm radio observations of
  the Galactic Centre region \cite{LaRosa00}. The Galactic Plane is
  oriented top-left to bottom-right in this image. The Galactic
  Centre is located inside the Sgr~A region.}
\end{figure}

In the first large-scale compilation of 90~cm radio observations
provided by La~Rosa, Kassim, and Lazio \cite{LaRosa00}, the central
few 100~pc region reveals a complicated morphology with various
objects, mostly supernova remnants, H II regions, and Giant Molecular
Clouds (see Fig. \ref{fig:LaRosa}). Thread-like filaments, notably the
GC radio arc, exhibit highly polarised radiation with no line
emission \cite{Yusef84}, and are therefore, amongst others, regions
with populations of non-thermal electrons, emitting synchrotron
radiation.  

The structure of molecular clouds in the region has been mapped
already in the 1970s using $^{12}\mathrm{CO}$ and $^{13}\mathrm{CO}$
lines (e.g. \cite{bania77, liszt77}). These measurements, however,
suffer from background and foreground
contamination from molecular clouds in the Galactic disk. In the
velocity range of interest for mapping the GC region, $|v|<30$~km/s,
the CS~(J=1-0) line is expected to be essentially free of such
contaminations. Albeit being less sensitive because of its
higher critical density, CS emission provides an efficient tracer of
the dense molecular clouds close to the GC. The most complete CS map
of the region is provided by measurements of the NRO radio telescope
\cite{cs} and yields a total mass in molecular clouds of $(2-5)\times
10^7 M_{\astrosun}$ in the inner 150~pc region. These clouds are a
potential target for cosmic rays accelerated within the region.

The radio view of the inner 50~pc region is dominated by the Sgr A
complex, with \astar\ at its centre. Along the line-of-sight of
\astar\ lies \aeast, at a projected distance of 2.5~pc from the GC, 
enclosing in projection Sgr~A~West, a three-armed structure which
spirals around \astar\ and exhibits a thermal spectrum.  

\aeast\ resembles a compact morphology because of the high
density of interstellar material ($\approx 10^3$~cm$^{-3}$) which prevents a fast
evolution of the forward shock. Furthermore, it interacts with a dense
molecular cloud (density $\approx 10^{5}$~cm$^{-3}$) on its eastern side.
Based on recent X-ray observations \cite{Maeda2002} and older radio
measurements \cite{Jones74}, \aeast\ is very likely the remnant of a
massive star which exploded about 10.000 years ago (SNR
000.0+00.0). An overabundance of heavy elements is
found, favouring a SN type II explosion of a 13-20~$M_{\astrosun}$ star. 
The X-ray emitting region of the remnant appears more
compact (2~pc radius) then at radio wavelengths (6-9~pc), and is caused
by the reverse shock that heats plasma in the inner parts of the
remnant.  

Observations in the near-infrared (NIR, e.g. \cite{Eisenhauer:2005cv}) have
been taken to precisely measure the orbits of young stars in the
direct (as close as 0.1'') vicinity of \astar. From these, the
distance of the solar system to the Galactic Centre, $d_{\mathrm{GC}}
= (7.62\pm 0.32)$~kpc, and the mass of the central compact object, 
$m_{\mathrm{A^*}} = (3.61\pm 0.32)\times 10^6~M_{\astrosun}$, can be
inferred \cite{Eisenhauer:2005cv}. The orbits are consistent with
Keplarian motion around a point mass centred on \astar. Furthermore,
VLBA measurements put constraining limits on the proper motion of
\astar, requiring an enclosed mass of at least $~4\times 10^5~M_{\astrosun}$
\cite{Reid2004}. At a wavelength of 7~mm, VLBI observations have
resolved the size of the radio emission region to $24\pm 2$
Schwarzschild radii \cite{Bower2004}. Combining these findings, there
is not much doubt that \astar\ can only be a supermassive black hole
(SMBH, see, e.g., \cite{Genzel:2007aa, Melia07} for recent reviews).
Its energy spectrum in the millimeter to IR domain is
characterised by a hard power-law with spectral index $\approx 0.3$,
a turn-over at about 1~GHz, followed by a cutoff at about $10^3$~GHz
\cite{Zylka1995}, explained as synchrotron radiation of relativistic
electrons (see e.g. \cite{Duschl1994, Melia2000}). 

While being relatively bright at radio frequencies, \astar\ is only a
faint X-ray source \cite{Skinner1987}, but shows bright outbursts on
time scales of a few minutes to several hours 
(see e.g. \cite{Baganoff2001, Porquet2003}). These short flare durations limit the
emission region to less than 10 Schwarzschild radii of the black hole,
where non-thermal processes near the event horizon might produce
relativistic electrons
(e.g. \cite{Markoff2001,Aharonian:2005ti,Liu2006}). Such models to a
certain extend predict flares in the NIR band. Such flares have been
observed \cite{Genzel2003}, but occur much more frequently
than at X-ray energies. In contrast to this, INTEGRAL observations in
the hard X-ray/soft \gr\ band show a faint, but steady emission from
the direction of the GC \cite{Belanger2006}.

In 1998, observations with the EGRET instrument onboard the Compton
Gamma-Ray Observatory provided a strong excess (3EG~J1746-2851) of
$>30$~MeV \grs\ on top of the expected Galactic diffuse
emission. Within an error circle of $0.2^\circ$, the position of this
excess is compatible with the position of \astar. However, although
not completely ruled out, the extension of the excess fits better a
picture where the emission is produced by several distributed
objects or diffuse interactions rather than by a single compact
object like \astar. Moreover, the energy output of 3EG~J1746
in the MeV-GeV range ($\approx 10^{37}$~erg~s$^{-1}$) exceeds by at
least an
order of magnitude the energy released close to \astar\ at any other
wavelength. In any case, due to the relatively poor angular
resolution of EGRET, source confusion hampers the interpretation of
the signal especially at low energies, where the EGRET point spread
function is worst. A follow-up analysis of the position of 3EG~J1746, 
using only events with energies $>1$~GeV, disfavours its
association with \astar\ at the 99.9\% CL
\cite{Hooper2002}. 

It has been argued that 3EG~J1746 may be
associated to the SNR \aeast, since its \gr\ spectrum is similar to
other SNRs detected by EGRET. One caveat is, however, that the \gr\
luminosity of 3EG~J1746 is two orders of magnitude larger than
what is found for other EGRET SNRs. Fatuzzo et al. \cite{Fatuzzo2003},
however, convincingly explain  
the high energy ($>100$~MeV) \gr\ emission from the source
as being produced in inelastic collisions of
shock-accelerated protons with the ambient medium, self-constitently
accounting for the $<100$~MeV and radio emission by bremsstrahlung
processes and synchrotron emission, respectively, of electrons
produced as secondaries in the decay of charged pions (via muon
decay). 

\section{Discovery of the GC in VHE \grs}

The recent detection of VHE \grs\ from the
direction of the Galactic Centre by several Imaging Atmospheric
Cherenkov Telescopes (IACTs)
\cite{Tsuchiya:2004wv,Kosack:2004ri,Aharonian:2004wa,Albert:2005kh}
has firmly established the existence of particle acceleration to
multi-TeV energies within the central few pc of our
galaxy. Prior attempts to detect VHE \grs\ from this region
with the HEGRA stereoscopic system were not successful, and a weak
flux upper limit of 8.7 times the flux of the Crab
nebula above 4.5~TeV was reported \cite{HEGRAGalacticPlane}. 

In 2004, the discovery of a VHE \gr\ signal from the direction of
the GC was almost simultaneously reported by the CANGAROO-II
\cite{Tsuchiya:2004wv} and Whipple \cite{Kosack:2004ri}
collaborations, with energy thresholds of 250~GeV and 2.8~TeV,
respectively. For both analyses the emission regions include the
position of \astar\ within the statistical errors, and are compatible
with a point-like origin. No hint for flux variability on
timescales of months or years is found. 

The CANGAROO-II telescope detects the source with a significance of
10~$\sigma$ above the background in 67~hours of observations.
The differential energy spectrum reported is very steep, $\propto
E^{-4.6\pm 0.5}$ \cite{Tsuchiya:2004wv}, with a flux normalisation at
1~TeV of about $2.7\times 10^{-12}$~cm$^{-2}$~s$^{-1}$~TeV$^{-1}$. 

In 26 hours of large zenith angle observations, Whipple detects the GC
with a marginal significance of 3.7~$\sigma$ above the background. The
integral \gr\ flux reported is $(1.6\pm 0.5_{\mathrm{stat}} \pm
0.3_{\mathrm{syst}})\times 10^{-12}$~cm$^{-2}$~s$^{-1}$ above 2.8~TeV,
roughly two orders of magnitude larger than the flux measured by
CANGAROO-II at these energies. 

The \hess\ telescope array observed the GC region first during the
commissioning of the 
partially incomplete array in June-August 2003. Only two of the four
telescopes were operational. During the first phase of measurements
(June/July 2003, 4.7 hours observing time) the telescopes were
triggered independently, and events were combined using GPS
timestamps. For the second observation campaign (July/August 2003, 11.8
hours observing time) a hardware stereo trigger was used, reducing the
energy threshold to 165~GeV for this data set. Also \hess\ reports the
detection of a point-like VHE \gr\ source (henceforth called \hgc)
coincident with \astar. Differential energy spectra have been produced
separately for the two data sets, best described by hard power-laws up
to the highest energies measured. 49 hours of observations with the
completed \hess\ array were carried out in 2004, yielding consistent
results. From a power-law fit of the 2004 data, a photon index of
$\Gamma=2.25\pm 
0.04_\mathrm{stat}\pm 0.10_\mathrm{syst}$ and an integral flux
above 1~TeV of $[1.87\pm 0.10_\mathrm{stat} \pm
0.30_\mathrm{syst}] \times 10^{-12}$~cm$^{-2}$~s$^{-1}$ is obtained
\cite{Aharonian:2006wh}. Recent MAGIC
observations of \hgc\ at large zenith angles in 2004 and 2005 
verify the hard spectrum found by \hess, with consistent flux
levels, and confirm the point-like and non-variable characteristics of
the source \cite{Albert:2005kh}. 

The \gr\ flux measured by \hess\ and MAGIC is a factor of three
lower than that provided by Whipple, and the hard spectral index is in
clear contradiction with the CANGAROO-II results. This either makes \hgc\
a rapidly varying \gr\ source (with the caveat being that none of the
experiments detected significant variability in its own data set),
or points to some hidden systematics in the analysis. Indeed, in a careful
reanalysis of the Whipple data \cite{Kosack2005} the flux level has
been corrected, and a differential energy spectrum which
matches the \hess\ and MAGIC spectra is obtained. Moreover,
observations with the CANGAROO-III array recently yielded a
differential energy spectrum consistent with the \hess\ and MAGIC
results \cite{Mizukami2008}.  

\begin{figure}
\includegraphics[width=0.48\textwidth]{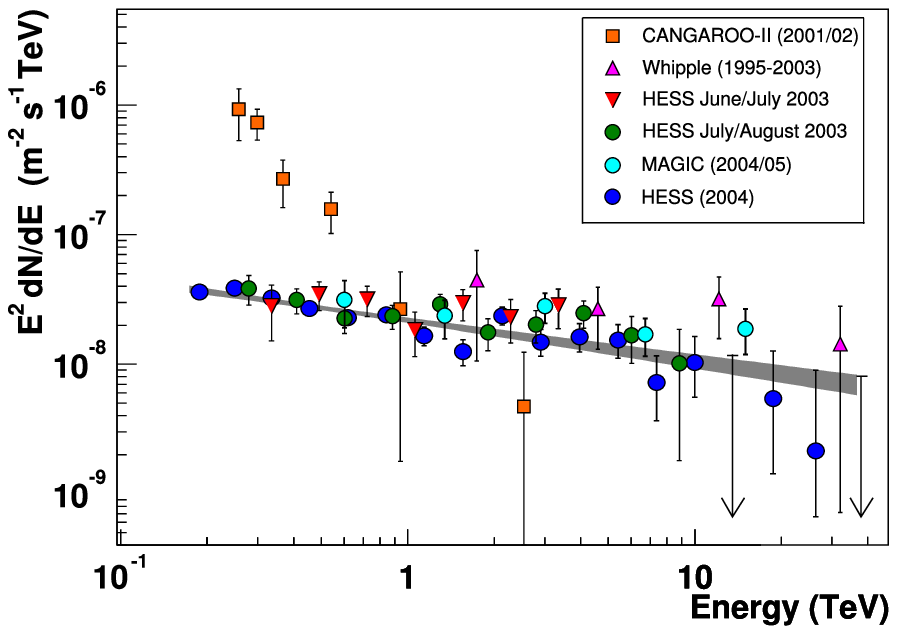} 
\label{fig:Spectra}
\caption{Compilation of spectral energy distributions ($E^2\times$~flux)
  of the GC source \hgc. Data points are
taken from \cite{Tsuchiya:2004wv,
  Aharonian:2004wa,Albert:2005kh,Aharonian:2006wh, Kosack2005}. 
The grey shaded band shows a power-law fit
$F(E)\propto E^{-\Gamma}$ to the \hess\ 2004 data
\cite{Aharonian:2006wh}. The recent CANGAROO-III result
\cite{Mizukami2008} shown at this conference is not yet included, nor
is the integrated flux measurement by Whipple \cite{Kosack:2004ri}. Note that for the
\hess\ 2004 result a contribution of 17\% of the total flux from diffuse
emission was subtracted first.} 
\end{figure}

Fig. \ref{fig:Spectra} shows a compilation of the at date available
VHE \gr\ flux measurements from the direction of \hgc,
indicating the recently achieved agreement between the experiments. 

\section{Today's VHE \gr\ view of the Galactic Centre Region}
\begin{figure*}
  \includegraphics[width=0.85\textwidth]{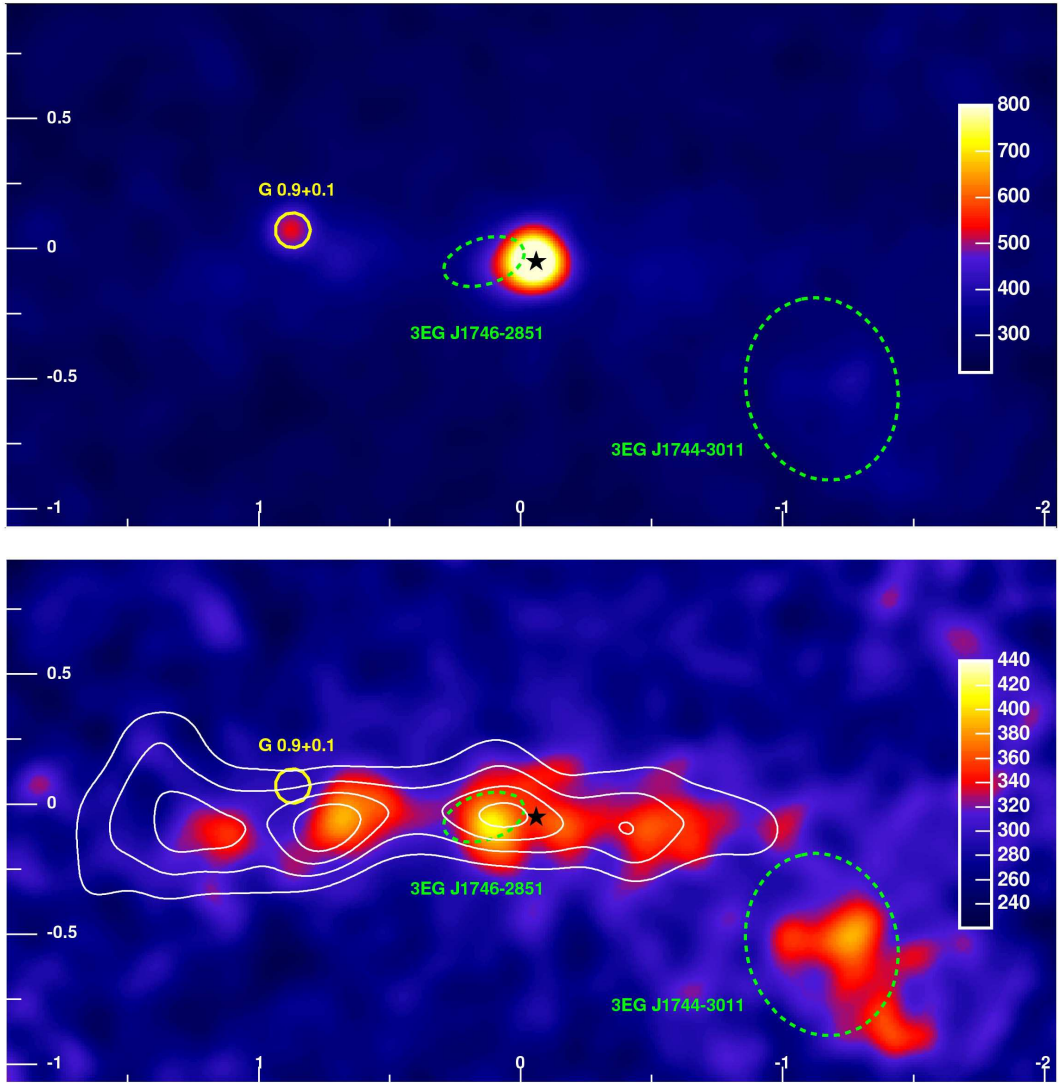}
  \caption{\hess\ VHE $\gamma$-ray images of the Galactic Centre region
    \cite{Aharonian:2006au}. Top: smoothed $\gamma$-ray count map without
    background subtraction showing the bright sources \hgc\ 
    (spatially coincident with the supermassive black hole Sgr~A*
    marked by a black star) and G0.9+0.1. Bottom: same map after
    subtraction of the two (assumed point-like) sources, showing an
    extended band of highly significant ($14.6~\sigma$) \gr\
    emission and the unidentified source HESS~J1745-303 (lower right corner). The white  
    contours show velocity integrated CS line emission \cite{cs}, smoothed to
    match the angular resolution of \hess. The positions of two
    unidentified EGRET sources are shown by green ellipses.} 
  \label{fig:GCDiffuse}
\end{figure*} 

Whilst the first detections of the GC were based on data sets of
limited statistics and/or high energy threshold, follow-up observations
with the completed \hess\ instrument provide much better sensitivity
and deliver the at date most sensitive VHE \gr\  images of the GC region
\cite{Aharonian:2006au}. Thanks to its large field-of-view \hess\ is
able to observe a region of $\approx$400~pc diameter (at an assumed
distance to the GC of 8~kpc) with a single pointing of
the instrument. The so-far published results are based
on a deep exposure of GC region with the full \hess\ array in 2004. In 49 
hours of quality-selected data, the detection of \hgc\ is confirmed 
with a high significance of about $38~\sigma$ above the background. 
In the \gr\ count map shown in Fig. \ref{fig:GCDiffuse} (top) a second
discrete source is visible $1^\circ$ away from \hgc, associated with
the SNR \gsnr.

While for previous VHE instruments sources like \gsnr\ were
close to the detection limit, the \hess\ data set enables the
search for fainter emission.  Subtracting the best-fit model for
point-like emission from the positions of \hgc\ and
G0.9+0.1 yields the sky map shown in the bottom part of
Fig. \ref{fig:GCDiffuse}. It reveals the presence of diffuse emission
along the Galactic Plane \cite{Aharonian:2006au}, as well as the
extended \gr\ source HESS~J1745-303 \cite{Aharonian:2006zz}, located
about $1.4^\circ$ south-west of \astar. HESS~J1745-303 belongs
to a rather long list of unidentified Galactic VHE \gr\ sources, for
which the lack of solid couterparts at other wavelengths renders
a firm identification difficult so far. A multi-wavelength
analysis including VLA, XMM and recent \hess\ data suggests that at
least parts of the emission of HESS~J1745-303 can be explained by
the shock wave of the SNR~359.1-00.5 (Fig. \ref{fig:LaRosa}) running
into a dense molecular cloud and by a pulsar wind nebula driven by the pulsar
PSR~B1742-30 \cite{Aharonian:2008aa}. An association of HESS
1745-303 and 3EG~J1744-3011 is difficult both in terms of the
energetics and possibly detected variability of the EGRET source.

\subsection{\gsnr: a Pulsar Wind Nebula  in VHE \grs}
\gsnr\ is a well-known composite SNR with a clear shell-like radio
morphology (see Fig. \ref{fig:LaRosa}). It consists of a radio
shell of 8' diameter and a bright compact core. Given the extension of
the shell and assuming a distance of 8~kpc, the supernova took
place a few 
thousand years ago. Although no pulsed emission has been detected from
the central core region, X-ray observations \cite{Gaensler2001,
  Porquet2003a} have identified it as a pulsar wind nebula (PWN) and
also found spectral softening away from the core of the nebula,
suggesting an electron population which cools due to synchrotron
radiation on its way outwards. No non-thermal X-ray emission
has been detected from the SNR shell. 

The \hess\ instrument has discovered \gr\ emission from the direction of
\gsnr\ with a significance of 13~$\sigma$ after 50 hours of
observations \cite{Aharonian:2005br}. The fact that \gsnr\ is not
detected by the MAGIC instrument is not in conflict with the \hess\
result, given the lower sensitivity of the MAGIC data set. The
morphology of the VHE source is 
compatible with a point-like excess, and an upper limit on the source
size of 1.3' at 95\%~CL is derived, excluding particle acceleration in
the SNR shell as the main source of the VHE \grs.
Instead the centroid 
of the \gr\ excess coincides within statistical errors with the
Chandra position of the PWN, making a PWN association compelling. 
The VHE \gr\ spectrum of \gsnr\ extends from 230~GeV to 6~TeV and is
best described by a straight power-law with a 
photon index of 2.40. The total power radiated in VHE \grs\ is
$2\times 10^{34}$~erg~s$^{-1}$, which can be easily accounted for in a
one-zone inverse Compton (IC) model yielding a
reasonable magnetic field strength ($6\mu$G) and a photon density of
5.7~eV~cm$^{-3}$ \cite{Aharonian:2005br}, somewhat smaller than the
conventional value used in GALPROP.

\subsection{\hgc: a Prime Example of an Unidentified \gr\ Source}
While the nature of the emission from \gsnr\ seems to be fairly settled,
this is certainly not true for the VHE emission from \hgc. Although
all IACTs which have observed the source have found it being positionally
coincident with 
the SMBH \astar, the actual mechanism that produces the emission is
still not identified. 
Besides a couple of different \astar-related
emission mechanisms proposed, there are at least two other objects in
direct vicinity of the SMBH which are convincing candidates
for producing to observed VHE \gr\ flux in parts or in
combination:

\begin{itemize}
\item Various models predict VHE \gr\ production near the
super-massive black hole (SMBH) itself \cite{Aharonian:2005ti} or in
termination shocks driven by a wind from the SMBH \cite{Atoyan2004}.
\item Annihilation of Dark Matter (DM) particles
clustering in a cusp around the SMBH could potentially produce VHE
emission \cite{Bergstroem2000}. 
\item The PWN \pwn, recently discovered in a deep 
Chandra exposure \cite{Wang:2005ya}, and only 8.7'' away from \astar\
in projection, may accelerate electrons to TeV energies. 
\item Finally, the SNR \aeast\ is a prime candidate counterpart,
given its non-thermal radio shell and the fact that SNRs are proven
sites of efficient particle acceleration to highest energies.
\end{itemize}

From the observer's standpoint,
an identification is particularly hampered by the relatively poor angular
resolution of current IACT installations which gives rise to source
confusion in this region. The point spread function (PSF) of the
\hess\ instrument is O(0.1$^\circ$), resulting in a relatively large
emission region (see Fig. \ref{fig:GCDiffuse}). 
Nevertheless can VHE \gr\ observations of
\hgc\ put constraints on counterparts and emission models in various
ways. The most sensitive data set is currently provided by the \hess\
collaboration. Without being in conflict with measurements at longer
wavelengths, models must explain the following properties of \hgc\
\cite{Aharonian:2006wh}:

\begin{itemize}
\item The centroid of \hgc\ is coincident within $7''\pm
  14_\mathrm{stat}'' \pm 28_\mathrm{syst}''$ with the
  position of \astar, and the intrinsic size of the source 
  amounts to less than 1.2' (95\% CL).
\item The spectrum measured between 160~GeV and 30~TeV can be
  characterised by a power-law with photon index $2.25\pm
  0.04_\mathrm{stat} \pm 0.10_\mathrm{syst}$. The integral flux
  above 1~TeV is $[1.87\pm 0.10_\mathrm{stat} \pm
0.30_\mathrm{syst}] \times 10^{-12}$~cm$^{-2}$~s$^{-1}$. This implies
a \gr\ luminosity of $10^{35}$~erg~s$^{-1}$ in the 1-10~TeV range. 
No hint for curvature is found. Assuming an exponential cutoff, a lower limit of
9~TeV (95\% CL) on the cutoff energy is derived.
\item There is no hint for significant flux variability on any
  timescale from minutes to years.
\end{itemize}

\subsubsection{The Case of \aeast}
The existence of synchrotron radiation, i.e. the presence of relativistic electrons,
and a large magnetic field ($\approx 2-4$~mG, determined from Zeeman
splitting of OH masers \cite{Yusef96}) make \aeast\ a compelling
candidate for \gr\ emission at VHE energies. 
In particular do the observated flux spectra at radio, X-ray, and \gr\
energies match a scenario in which protons have been shock-accelerated to at
least 100~GeV of energy \cite{Fatuzzo2003} (see above). Adopting a
4~mG magnetic field, Crocker et al. \cite{Crocker2005} estimate the
maximum proton energy achievable in the \aeast\ blast wave to be
$10^{19}$~eV. The fact that none of the IACTs reports flux variability
from \hgc\ certainly fits into this scenario.

The shell of \aeast\ partially surrounds \astar\ in projection; its
emission maximum in 90~cm radio is 1.5' (or about 3.5~pc)
away from \astar. Due to the systematic
error of 28'' on the position of the \hgc\ centroid from uncertainties in
the absolute pointing of the \hess\ telescopes, the position of the
VHE emission is marginally consistent with the \aeast\ radio maximum. 

\begin{figure}[htbp]
  \includegraphics[width=0.48\textwidth]{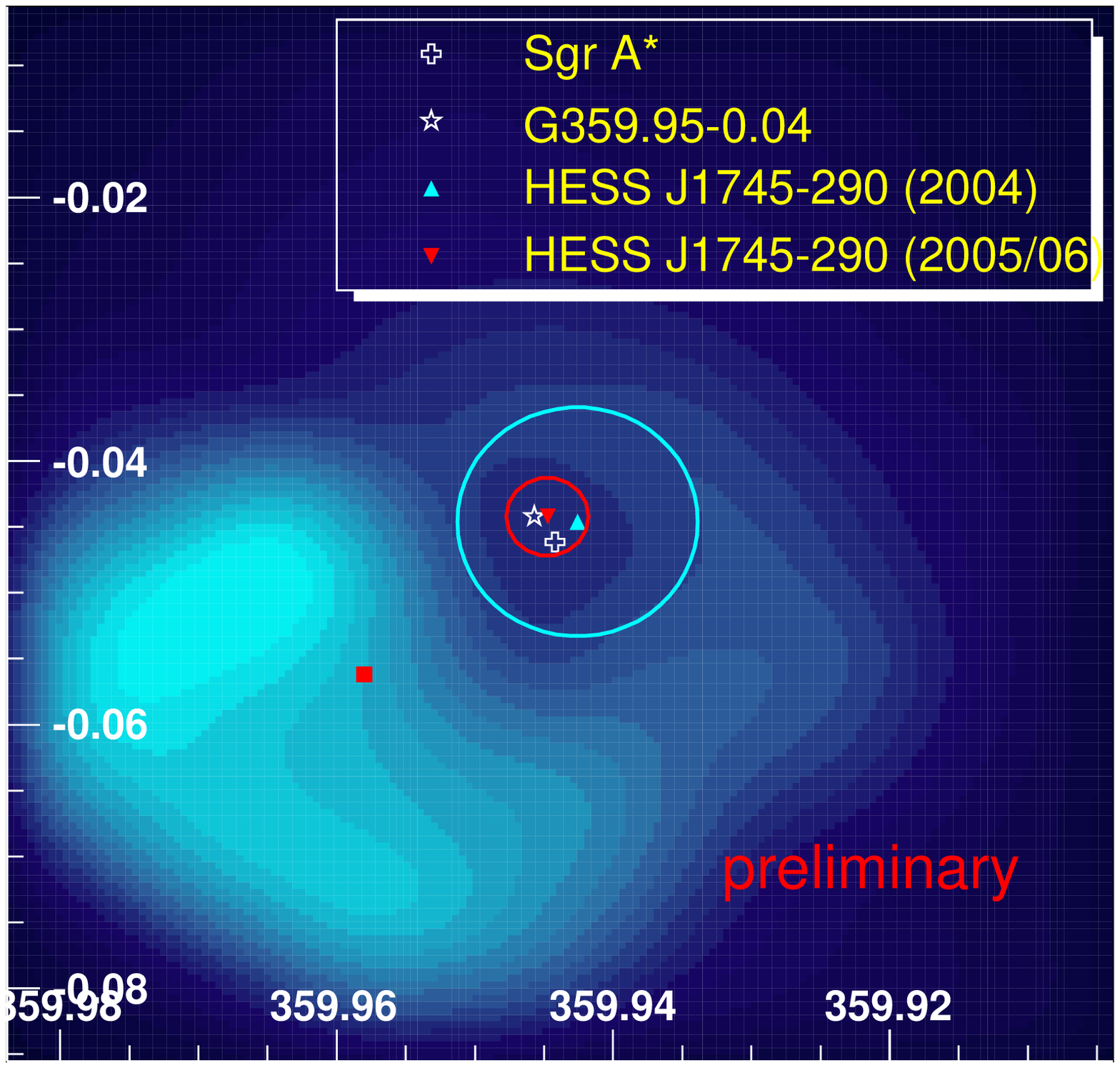}
  \caption{Best fit positions of \hgc\ on top of a smoothed 90~cm VLA
    radio image of SNR \aeast\ in Galactic coordinates. The
    position of \astar\ and \pwn\ are marked with a cross and a star,
    respectively. The red triangle and red circle mark the preliminary best fit
    position and total error of the improved position measurement 
    of \hgc. The blue triangle and circle show the results obtained in
    \cite{Aharonian:2006wh}. Figure taken from \cite{vanEldik:2007icrc}.}
  \label{fig:GCPosition}
\end{figure}

Recent progress (although subject to final checks, \cite{vanEldik:2007icrc})
in understanding and compensating the pointing systematics of the
\hess\ array has led to a reduced systematic error of 8.5'' 
\cite{vanEldik:2007icrc}. 
Fig. \ref{fig:GCPosition} shows the improved \hess\ position
measurement, based on 73~hours of observations, on 
top of a 90~cm VLA radio image of \aeast. The best fit position is
coincident within $7.3'' \pm 8.7''_\mathrm{stat} \pm8.5''_\mathrm{syst}$ with
\astar\ and effectively rules out \aeast\ as the dominant source of the
VHE emission.

\subsubsection{\hgc: a Pulsar Wind Nebula?}
The recent detection of the PWN \pwn\ in a deep Chandra
exposure of the GC region \cite{Wang:2005ya} very much complicates
the counterpart search for \hgc. \pwn\ is located only $8.7''$ in
projection (or 0.3~pc) away from \astar, rendering a discrimination of the two by
position measurements impossible (cf. Fig. \ref{fig:GCPosition}).  
\pwn\ is rather faint at X-ray energies, with an implied luminosity
of $10^{34}$~erg~s$^{-1}$ in the 2-10~keV band
\cite{Wang:2005ya}, yet about four times brighter than \astar. It
shows a cometary shape and exhibits a hard 
and non-thermal spectrum which gradually softens when going away from
the ``head'' of the PWN, where the yet undiscovered pulsar is believed
to be located. No radio counterpart of the PWN is found.

Numerical calculations show that, despite its faint X-ray flux, a
population of non-thermal electrons can naturally explain both the
X-ray emission of \pwn\ and the VHE \gr\ emission 
of \hgc\ \cite{Hinton:2006zk}. Compared to other locations in the
galactic disk, where many VHE \gr\ sources have been found associated
to PWNe, the Galactic Centre region is, however, special because of its dense
radiation fields. The fact that the X-ray spectra steepen rather than
harden the further one gets away from the pulsar position is an indication that the
TeV electrons are cooled by synchrotron radiation rather than by
IC processes in the Klein-Nishina regime, putting a lower
limit of $\approx 100 \mathrm{\mu G}$ on the value of the magnetic
field for typical Galactic Centre radiation fields
\cite{Hinton:2006zk}. It is predominantly the far-IR component
of the radiation field that TeV electrons upscatter to TeV energies,
providing roughly an order of magnitude larger luminosity in the
1-10~TeV \gr\ band than in the 2-10~keV X-ray domain.

\subsubsection{Emission models involving \astar}
The low bolometric luminosity ($< 10^{-8} L_{\mathrm{Edd}}$ in the
range from millimeter to optical wavelenghts) renders \astar\ an
unusually quiet representative of galactic nuclei. At the same time,
this property makes the immediate vicinity of the SMBH transparent for
VHE \grs. Aharonian~\&~Neronov \cite{Aharonian:2005ti} show that the
absence of dense IR radiation fields enables photons with an
energy of up to several TeV to escape almost unabsorbed from regions
as close as several Schwarzschild radii from the centre of the SMBH.
Therefore, VHE \gr\ emission produced close to the event
horizon of \astar\ provides a unique opportunity to study particle
acceleration and radiation in the vicinity of a black hole. 

There are several possibilities to produce the observed VHE \gr\ flux, 
depending on the type of particles accelerated, the model of
acceleration, and finally the  
interaction of the accelerated particles with the ambient magnetic
field or matter. Common scenarios, which do not contradict the
emission at longer wavelenghts, include \cite{Aharonian:2005ti}: 

\begin{itemize}
\item Synchrotron radiation of ultra-relativistic protons. In the
  strongly magnetised environment of a SMBH with magnetic field
  strengths as large as $10^4$~G, protons can be accelerated to
  energies of up to $10^{18}$~eV. However, the synchrotron spectrum
  extends only to roughly 300~GeV, and thus cannot account for the
  multi-TeV radiation seen by \hess. Curvature radiation of protons
  can in principle extend the \gr\ spectrum 10~TeV, but only at the
  expense of very large magnetic fields ($10^6$~G), for which the
  source is opaque for \grs\ due to $e^+e^-$ pair production.
\item Photo-Meson interactions. Despite its low bolometric luminosity,
  the IR radiation fields in the vicinity of \astar\ appear dense
  enough to produce a sizable number of VHE \grs\ in the interactions
  of the accelerated protons with IR photons. The required
  power to meet the 
  luminosity in VHE \grs\ is $10^{38}$~erg~s$^{-1}$, well below the
  Eddington luminosity of a $3\times 10^6 M_{\astrosun}$ BH.
\item Proton-Proton interactions. VHE \gr\ production by interactions
  of accelerated protons with the ambient plasma requires an
  acceleration power of $10^{39}$~erg~s$^{-1}$, but at the same time
  only $\geq 10^{13}$~eV protons are required to produce TeV
  radiation. In this case, possible acceleration sites include a
  strong electric field close to the event horizon or strong shocks in
  the accretion disk. This scenario in particular predicts correlated flux
  variability at VHE, X-ray, and IR wavelengths.
\item Inverse Compton radiation of electrons. Compared to the
  aforementioned proton scenarios, electrons provide a much more
  efficient way to convert energy into radiation. To accelerate
  electrons to multi-TeV energies, however, a well-ordered magnetic
  or electric field is necessary to prevent radiation losses during
  acceleration. Such properties are e.g. provided by the
  rotation-induced electric fields near the black hole. \grs\ at the
  highest energies ($> 100$~TeV), however, cannot escape the source because of
  efficient interaction with the IR radiation field, but 
  contribute to the spectrum with sub-100~TeV photons.
\end{itemize}

While some of the above mentioned scenarios suggest correlated
multi-wavelength variability, non-observation of variability does not
striktly rule out acceleration close to the black hole. Moreover,
there are models which explain the absence of VHE variability by
diffusion of protons away 
from the acceleration region into the neighbourhood of
\astar\ and subsequent interaction with the ambient medium \cite{Liu2006a,
  Liu2006}. On the other hand, the detection of variability in the
VHE data would immediately point to particle acceleration near the
SMBH. The most convincing signature would be the detection of
correlated flaring in X-rays (or NIR) and VHE \grs. Such
searches have been carried out \cite{vivier07,hinton07}. No evidence
of flaring or quasi-periodic oscillations has been found.
In particular, in a coordinated multi-wavelength campaign both Chandra
and \hess\ observed the GC region, when a major (factor 9 increase)
X-ray outburst was detected. During this 13-minutes flare the VHE \gr\
flux stayed constant within errors, and a 99\% CL upper limit on a doubling 
of the VHE flux is derived \cite{hinton07}. 

\subsubsection{Dark Matter Annihilation near the GC?}
Besides being of astrophysical origin, the observed TeV flux could
potentially stem from annihilation of dark-matter particles,
which are believed to cluster in a compact cusp around Sgr A*
\cite{Bergstroem2000}. Halo density profiles are believed to scale
with the radius $r$ like $r^{-\alpha}$, with $\alpha$ between 1
\cite{Navarro1997} and 1.5 \cite{Moore1999} in the most common
models. The fact that \hgc\ is point-like (after having accounted for
the underlying diffuse emission) translates into $\alpha>1.2$,
i.e. a cuspy halo is favoured by the observations. 

\begin{figure}
  \centering
  \includegraphics[width=0.46\textwidth]{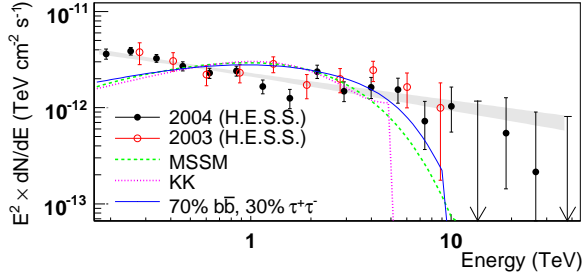}
  \caption{Spectral energy density of
    \hgc. The shaded band shows the power law fit
    $dN/dE\sim E^{-2.25}$ to the 2004 data points
    (\cite{Aharonian:2006wh}, see also Fig. \ref{fig:Spectra}. The
    curves show  
    typical spectra of \grs\ from the annihilation of 14~TeV
    MSSM neutralinos (green), of 5~TeV Kaluza-Klein particles
    (purple), and of a 10~TeV DM particle decaying into 30\%
    $\tau^+\tau^-$ and 70\% $b\bar{b}$ (blue). Figure reproduced from 
    \cite{Aharonian:2006wh}.}  
  \label{fig:GCSpectrum}
\end{figure}

Predicted energy spectra for \grs\ produced in cascade decays
of DM particles such as MSSM neutralinos or Kaluza-Klein particles can be
compared to the VHE observations. These spectra are usually curved
both at high energies -- for reasons of energy conservation --, and low
energies, in clear disagreement with the observations
(Fig. \ref{fig:GCSpectrum}, see also
\cite{Aharonian:2006wh}). Furthermore, unusually large DM particle  
masses have to be assumed to account for the fact that the \gr\
spectrum extends up to 10~TeV. 

The observed \gr\ emission is therefore not compatible with 
being dominantly produced in the framework of the most common DM
scenarios. As a consequence, the bulk of the 
observed \gr\ excess is probably of astro- rather than of particle
physics origin. However, an O(10\%) admixture of \grs\ from DM 
annihilations in the signal from the GC cannot be ruled out. 
Assuming a NFW-type \cite{Navarro1997} halo profile, 99\% CL upper
limits on the velocity-weighted annihilation cross section $<\sigma
v>$ are at least two orders of magnitude above theoretical
expectations \cite{Aharonian:2006wh}, and thus are not able to put
constraints on current model predictions. 

\subsection{Diffuse \gr\ Emission}

The diffuse emission (Fig. \ref{fig:GCDiffuse} bottom) spans in a region of roughly  
$2^\circ$ in galactic longitude ($l$) with an rms width of about $0.2^\circ$
in galactic latitude ($b$). The reconstructed $\gamma$-ray spectrum
integrated within $|l|\leq 0.8$ and $|b|\leq 0.3$ is well-described by
a power law with photon index $\Gamma=2.29$ \cite{Aharonian:2006au},
in agreement with the index observed for \hgc. Assuming that the emission is
produced in the Galactic Centre region at 8~kpc distance from the observer,
the latitude extension translates into a scale of about
30~pc. This is very similar to the extent of giant molecular clouds in
this region \cite{cs}. Indeed, at least for $|l|\leq 1^\circ$, there is a
strong correlation between the morphology of the observed
\grs\ and the density of molecular clouds as traced by the CS 
emission observed with NRO (\cite{cs}, Fig. \ref{fig:GCDiffuse} bottom). 
This is the first time such correlation is seen, and is a strong
indication for the presence of an accelerator of (hadronic) cosmic
rays in the Galactic Centre region, since the energetic hadrons would
interact with the material in the clouds, giving rise to the observed
\gr\ flux via $\pi^0\to\gamma\gamma$ decays. The idea of acceleration
in the GC region is further supported by the fact that the
measured \gr\ flux is both larger and harder than expected in a
scenario where the molecular material is bathened in a sea of galactic
cosmic rays, with similar properties as measured in our solar
neighbourhood, only. The energy
necessary to fill the entire region with cosmic rays can be estimated
from the measured \gr\ flux (extrapolated to 1~GeV) and amounts to
$10^{50}$~erg. This number is close to the energy believed to be
transferred into cosmic rays in a typical galactic supernova. 
A distribution of electron accelerators, such as PWNe, that cluster
similarly to the gas distribution, has also been discussed
(e.g. \cite{Aharonian:2006au}). Given the O(100$\mathrm{\mu}$G)
magnetic fields in the region, electrons of several TeV energy would,
however, rapidly cool via synchrotron radiation, such that their VHE
\gr\ emission would appear point-like in the \hess\ data. 

In the context of identifying the accelerator, the fact
that no emission is seen farther than $|l|\approx 1^\circ$ might be
particularly important. In their discovery paper
\cite{Aharonian:2006au}, the \hess\ collaboration came up with the
rather simple, yet convincing explanation that the cosmic rays may
have been accelerated in a rather young source near the very centre of
the galaxy, having subsequently undergone diffusion away from the
accelerator into the surrounding medium. Assuming a typical diffusion
coefficient of $10^{30}$~cm$^2$~s$^{-1}$, or 3~kpc$^2$~Myr$^{-1}$, for
TeV protons in the Galactic disk, a source age of about $10^4$ years
can reproduce the observed \gr\ flux distribution
\cite{Aharonian:2006au}, and in particular the lack of emission beyond
$1^\circ$ distance from the centre.  

\begin{figure}
  \centering
  \includegraphics[width=0.46\textwidth]{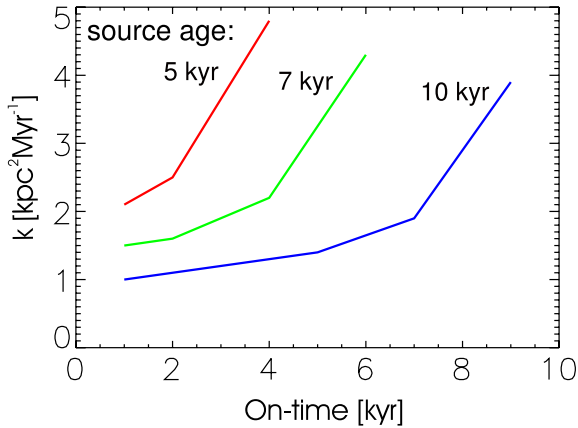}
  \caption{Best-fit diffusion coefficients (see
    text) for CR
    diffusion away from a central source into the GC region. Diffusion
    coefficients are given as a function of source on-time, for three
    different source ages. Reproduced from \cite{Buesching2007}.}
  \label{fig:DiffusionCoeff}
\end{figure}

B\"usching et al. \cite{Buesching2007} follow a
similar idea. Starting from a source of non-thermal protons at the GC
and the known distribution of molecular material,
the authors model the \gr\ flux from the region in a time dependent
diffusion picture. Neglecting a possible energy dependence of the
diffusion process (suggested by the fact that the \hess\ data are not
sensitive enough yet to measure such an effect), they use $\chi^2$
minimisation to find the diffusion coefficient for which the \hess\
results are matched best, for a variety of source
ages and source on-times (Fig. \ref{fig:DiffusionCoeff}), resulting in
diffusion coefficients in the range of the one assumed in
\cite{Aharonian:2006au}. In a similar approach, Dimitrakoudis et
al. \cite{Dimitrakoudis2008} obtain a best-fit diffusion coefficient
of 3~kpc~Myr$^{-1}$, also close to \cite{Aharonian:2006au}. Scaling the
diffusion coefficient $k$ with the cosmic ray rigidity $\zeta$, $k=k_0
(\zeta/\zeta_0)^{0.6}$, $\zeta_0 = 1$~GV$/c$, B\"usching et al. find a
value of $k_0$ which is significantly smaller than the local value,
suggesting enhanced turbulence and larger magnetic fields than in the
solar neighbourhood. Uncertainties in the derived diffusion
coefficients of up to 50\% arise from uncertainties in the molecular
gas distribution and from the contribution of galactic cosmic rays to
the overall \gr\ flux. 

In a separate paper \cite{Buesching2008}, B\"usching et al. explain 
both the diffuse emission and the point-source \hgc\ within a single
model in which the cosmic rays responsible for the diffuse emission were
accelerated in the shock wave of the SNR \aeast\ 5-10~kyr ago, but
acceleration stopped well before the present time. When, however, the
shock wave of \aeast\ collided with \astar, particle acceleration near
the SMBH was initiated, leading to the observed VHE \gr\ emission from
\hgc. Assuming that the diffusion coefficient found for the diffuse
emission is also valid close to the SMBH, this last round of particle
acceleration can only have happened in the recent past
(O(100)~yr) to be consistent with the point-like morphology of \hgc. 

It should, however, be noted that there are other processes which can
explain the emission from \hgc\ (see above). Furthermore, recent
simulations seem to indicate that the diffuse emission might be better
explained by inter-cloud acceleration of cosmic rays via the Fermi-II
process \cite{Wommer2008}. More sensitive observations are needed to
ultimately prove which of the discussed scenarios of the VHE \gr\ view
of the GC is correct.

\section{The Role of CTA}
Despite the exciting progress in
recent years, a robust understanding of the GC VHE \gr\ sky needs
a more refined data set than currently available. Significant
progress in the identification of the VHE sources and the physics
processes involved requires an instrument 
with better sensitivity, wider energy coverage, and, possibly, improved
angular resolution. Probing the GC
region with instruments like CTA \cite{Hermann2007} or AGIS will
answer many of the open questions within a reasonable amount of
observing time. 

For the following order-of-magnitude estimations we
assume a CTA-like array of IACTs with a core sensitivity of
O(1~mCrab), 4 orders of magnitude energy coverage (10 GeV -- 100 TeV),
and an angular resolution of 0.02$^\circ$ per event.

\subsection{Angular resolution}
As discussed above, one way of identifying \hgc\ is to search for
plausible candidate counterparts within the error circle of the
emission centroid. Since the GC is a densely packed region, naively, a
good angular resolution $\theta$ is important. Since the statistical 
error of the centroid position scales like $\theta/\sqrt{N_\gamma}$,
and therefore linearly 
with sensitivity, a factor of 50 improvement in the statistical error
of the centroid position over \hess\ (about 6'' per 
axis for 73 hours of exposure, \cite{vanEldik:2007icrc}) is expected. On the
other hand, the systematic pointing error of the \hess\ telescopes is
about 6'' per axis \cite{vanEldik:2007icrc}, probably close to the limit of
what can be achieved with future instruments. In the special case of \hgc,
subtracting the underlying (asymmetric) diffuse emission imposes
additional systematic uncertainties, which for \hess\ are of the order
of 1''. As a consequence, improved angular resolution is not of much help
what regards a measurement of \hgc's position.

\begin{figure*}
\centering
\begin{minipage}[b]{0.49\textwidth}
\includegraphics[width=\textwidth]{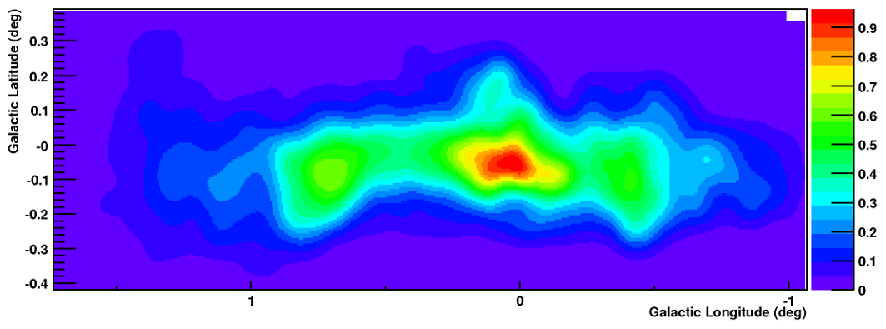}
\end{minipage}
\begin{minipage}[b]{0.49\textwidth}
\includegraphics[width=\textwidth]{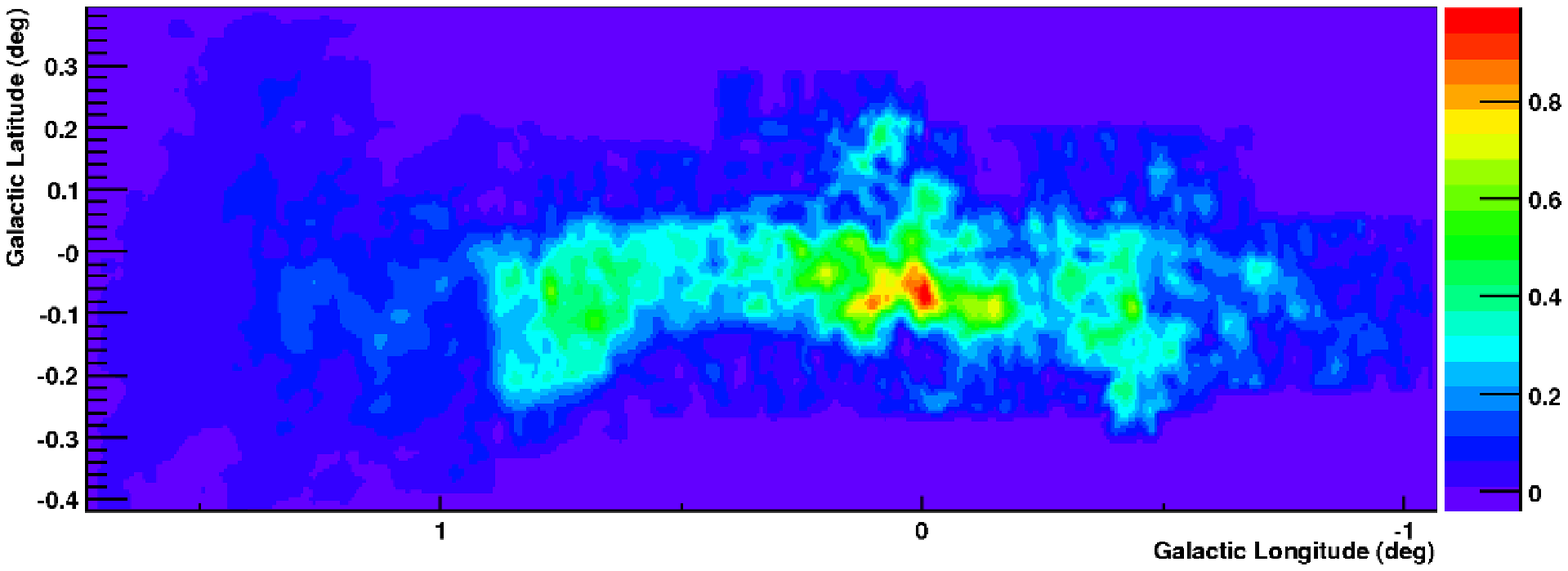}
\end{minipage} 
\label{fig:DiffuseCTA}
\caption{VHE \gr\ images of the GC region obtained from simulations. Cosmic rays 
produced at the Galactic Centre diffuse away from the centre and eventually interact
with molecular material (traced by CS emission \cite{cs}) in the surrounding and produce 
\grs. The diffusion coefficient assumed is $D=3$~kpc$^2$~Myr$^{-1}$, and the cosmic rays
have been injected $10^4$ years ago, matching the \hess\ measurements
of the diffuse emission \cite{Aharonian:2006au}. \emph{Left:} Expected
smoothed \gr\ map for an instrument with \hess-like angular resolution
($\theta_{68}=0.07^\circ$, taken from \cite{vanEldik:2007icrc}), and
\emph{Right:} for a CTA-like system with an assumed angular resolution
of $\theta_{68}=0.02^\circ$. Note that statistical fluctuations due to
possibly limited observing time are not taken into account.}  
\end{figure*}

On the other hand, superior angular resolution would help in
understanding the properties of the diffuse emission. It would allow
to probe the region with a few pc binning and test the \gr-cloud
correlation in much more detail than currently possible. 
Fig. \ref{fig:DiffuseCTA} sketches the improvement in angular resolution
over \hess\ for a CTA-like instrument with an angular resolution
of $0.02^\circ$ per reconstructed \gr. Shown are \gr\ maps expected
from cosmic rays interacing with the molecular material in the GC region. 
The cosmic rays diffuse away from the Galactic Centre, their assumed origin of production. 
The improvement in the quality of the data is clearly visible. 
With such an improvement one might in the not too distant future also get a handle on the
possible existence of electron accelerators along the Galactic Plane, which might or might not be
responsible for the observed \gr\ emission in parts or in total.

\subsection{Sensitivity}
An unambiguous proof that the VHE \gr\ emission from \hgc\ is associated
with \astar\ would be the observation of correlated \gr/X-ray (or
IR) variability. With the assumed CTA sensitivity similar X-ray flare
events like the one discussed above (factor 9 increase over the
quiescent level) would test a level as low as 10\% of the 
quiescent state \gr\ flux.

To test models of cosmic ray/electron propagation through the central
region of the galaxy and to study the penetration of molecular clouds
by cosmic rays, energy spectra have to be provided of the diffuse \gr\
emission in small regions of a~few~$10$~pc~$\times$~a~few~$10$~pc
only. With these at hand, energy-dependent diffusion processes could be
studied in great detail. For a simple power-law fit, the statistical
error on the spectral index scales linearly with
sensitivity. Therefore, because of its enhanced sensitivity, a
spectrum measured by CTA in a 
$0.1^\circ\times 0.1^\circ$ portion of the sky (in the \hess\ energy
range) will reach comparable statistical accuracy in the spectral
index as \hess\ does in a $1^\circ\times 1^\circ$ sky area
(e.g. $\Delta\Gamma_{stat} = 0.07$ for 50 hours of observations
of the GC diffuse emission \cite{Aharonian:2006au}). 

\subsection{Energy coverage}
The energy spectrum of \hgc\ covers an energy range of 160~GeV --
30~TeV and is well fitted by a straight powerlaw (see above). For the
most likely counterparts of the VHE emission, \astar\ and \pwn,
emission models fitting the combined spectral energy distributions
have been presented by various authors (e.g. \cite{Hinton:2006zk},
\cite{Aharonian:2005ti}, among others). While most models can
satisfactorily fit the \hess\ data 
points, they do substantially differ at energies $< 100$~GeV. CTA
energy coverage down to 10~GeV would constrain some of the
models, and therefore help to identify the source of the \grs\ and the
underlying physical acceleration and radiation processes.

\section{Conclusions}
Less than five years after the discovery of VHE \gr\ emission from the
direction of the GC, observations with Imaging Atmospheric Cherenkov
Telescopes provide a very sensitive view of this interesting region.
With the recent data from the \hess\ instrument, a rich VHE \gr\
morphology becomes evident, giving strong evidence for the existence of
a cosmic ray accelerator within the central 10~pc of the Milky Way.

An intense \gr\ point source is found coincident within errors with the
position of \astar. Source confusion near the GC make a solid
identification difficult, given the -- compared to X-ray 
satellites or IR observatories -- moderate angular
resolution of current IACTs. Recent progress in
improving on the systematic and statical errors of the centroid of
HESS~J1745-290 effectively excludes the SNR \aeast\ as the dominant
source of the \gr\ emission. A major contribution from the
annihilation of DM particles can also be excluded.

Future observations with even more
sensitive instruments such as CTA will significantly advance
our knowledge about the GC region at VHE energies. The
recently launched Fermi satellite will extend the energy range
down to about 100~MeV, such that unbroken sensitivity coverage will
be provided over 6 orders of magnitude in energy.



\begin{theacknowledgments}
The author would like to thank the organisers for
having invited him to present this overview at the symposium.
\end{theacknowledgments}



\bibliographystyle{aipproc}   

\bibliography{vaneldik_GC_review}

\IfFileExists{\jobname.bbl}{}
 {\typeout{}
  \typeout{******************************************}
  \typeout{** Please run "bibtex \jobname" to optain}
  \typeout{** the bibliography and then re-run LaTeX}
  \typeout{** twice to fix the references!}
  \typeout{******************************************}
  \typeout{}
 }

\end{document}